# A Universe Programmed with Strings of Qubits
By Philip Gibbs

**Abstract**
It has been suggested that reality works like a quantum computer, but such claims are just words if they are not backed up by sound mathematics. In pursuit of the fundamental equations I look to string theory where physicists led by Mike Duff have noticed useful connections between the quantum gravity of black holes and quantum information theory. By building on my earlier work on universal symmetry in string theory and using links between elliptic curves and hyperdeterminants, I find intriguing clues that these connections may be deep as well as useful. Ultimately any theory of the foundations of physics must explain why there are four forces and three generations of fermions. In string theory this would be a consequence of the choice of vacua. If a consistent formulation of string theory constructed from quantum bits can be found, it may be possible to understand the vast landscape of possibilities better and reverse engineer the program that codes our universe.

## Reality as a Quantum Computer

Ever since Shannon linked physical entropy to information [1], the idea that the universe is some kind of information processing system has gained widespread appeal. Wheeler used the enduring phrase "It from bit" [2] to express the way that existence itself might emerge from the concept of digital information, but the idea that the universe is a computer program running in another world goes back further. Computer pioneer Edward Fredkin has been especially influential with his theory of "digital physics" in which the universe is a cellular automaton [3]. Fredkin worked with Richard Feynman and inspired his seminal work on the theory of quantum computation. Now, in a modern version of digital physics, Seth Lloyd proposes that the universe runs as a quantum computer [4], but how viable is such a theory?

In recent times the principle of holography [5] has been used to explain why information is not lost in a black hole. The principle requires that the information bits describing the state of the black hole are stored on the black hole horizon, one for every Planck unit of area so that it can be recovered as the black hole evaporates due to Hawking radiation. This compelling and successful argument places discrete information at a fundamental level and forces us to consider the possibility that space and time themselves are quantised in discrete units. It is no coincidence that one of the earliest papers written on the holographic principle by Gerard 't Hooft used a model of a quantised cellular automaton [6] : The idea came to 't Hooft after discussions with Fredkin at Caltech. Now it is being found that the mathematics of quantum computing with quantum bits (known as qubits) is also useful to express the symmetries of black holes with duality in string theory [7,8]. So far this relates to just three or four qubits at a time, but could it be a further sign of a deeper truth about the universe? Could it be that the universe itself runs like a quantum computer processing qubits of information and calculating the evolution of space and matter?

I will present the case that this is indeed correct. In this view, continuous space and time are not fundamental. They emerge from the interactions of qubits. Our very existence depends on the fact that this quantum computer runs in a way that is currently beyond our technology. Decoherence which would blur our image of the world is controlled by algorithms we do not yet understand, perhaps by the action of quantum error correcting codes which appear naturally in the mathematics. Admittedly it is a speculative theory, but if it is true, it may help us to understand the principles we need to build the next generation of powerful computing technology. This is a possibility that cannot be ignored.

## A Discrete Universe?

The concept of discreteness in nature has a long pedigree. In ancient Greece Democritus famously invented the indivisible atom as an explanation for properties of matter. In the philosophy of that time, the distinction between matter and the space it inhabits was not often expressed. When the Arabic world of the Middle Ages adopted Greek science they interpreted the discreteness of nature as a property of space itself and encapsulated the idea into their geometric artwork to decorate buildings with tessellated figures. Later in the work of European scientists and mathematicians such as Newton, a clear line was drawn to differentiate between matter and the arena of space within which it acts. After the nineteenth century revival of the atomistic theory, the standard picture consisted of discrete particles of matter inhabiting a continuous space. Despite the rise of quantum theory which blurred the distinction between wave and particle, it is still the most widely held view that space and time are continuous while matter is discrete. Is this just a product of years of education which brainwashes us to place objects using coordinates on a Cartesian plane using continuous coordinates, or is it really the most natural picture of the world? More importantly, what does empirical observation tell us?

From theories of quantum gravity we have learnt that spacetime is unlikely to be described by the smooth manifold structures we use in classical physics. Gravity is not just another force. It is a feature of spacetime described by curvature tensors at the classical level. To quantise gravity we need to quantise spacetime itself in some sense. Measurements of spacetime at extremely small scales are likely to be limited by quantum effects. The result is that we do not expect to be able to measure details smaller than the Planck distance of around $10^{-35}$ meters or durations shorter than the Planck time of $10^{-43}$ seconds. This presents a very different picture from classical physics where there is no limit to how small a wavelength can be. How is this limitation realised? Is spacetime made of discrete blocks at the Planck scale?

Luckily we have one piece of observational input that can help resolve this question. On $10^{th}$ May 2010 a handful of high energy photons reached Earth from a gamma ray burst many billions of light years away. One photon with an energy of 31 GeV hit the detector of the Fermi Gamma Ray Telescope in Earth orbit, and it did so to within a second of other lower energy photons [9]. This is highly significant because many theories of discrete spacetime predict slightly different speeds for photons of differing energy. For the photons to arrive so close after such a long time the difference in speed must be very small indeed, smaller than expected from these theories. The inescapable conclusion is that spacetime behaves like a smooth and continuous Lorentz invariant spacetime even well below the Planck distance scale. If we are to replace spacetime with something else at very high energies, we must do so in a way that preserves its symmetries as if it is continuous.

## Emergent Spacetime

It is a widely held view that spacetime is not fundamental, but instead emerges from a description of interactions between particles that remains somehow free from the constraints of placement in a background spacetime. This is a natural idea that dates back to the time of Newton who first set out the traditional picture of particles moving in such a background. His rival Leibniz was one of the first to challenge this view. Like other philosophers who followed, he preferred to base his conceptualisation of the world primarily on the material objects we experience most directly rather than the space and time which seems more distant from our senses.

It is possible to imagine a model of reality in which the vacuum emerges from a sea of interacting spacetime atoms, just as a continuous fluid or solid is made of molecules, but in such a picture the discrete atoms would affect the passage of light waves in a way we could detect. Just as we can probe the structure of a crystal using diffraction, we should be able to probe the structure of spacetime using the passage of light or other waves. The Fermi observations show that we may have to accept that such probes won't work. This means that the process behind the emergence of spacetime must be something quite different. Luckily we can use mathematics to find out how this can happen.

**Four Qubits**

Luckily we can use mathematics to find out how spacetime can emerge with perfect local Lorentz symmetry. To see this, let us start from the idea that physics emerges from a system of interacting qubits. These are quantum systems described by a supposition of two spin states

$$\psi = \psi_0|0\rangle + \psi_1|1\rangle$$

Such states in relativistic physics have spin equal to $\frac{1}{2}$. To generate geometry from these states we need to use four qubits $\psi^{(1)}, \psi^{(2)}, \psi^{(3)}, \psi^{(4)}$, so that the four spin $\frac{1}{2}$ components can combine to give a spin 2 graviton. An interaction term would take a form like this

$$S_I = G^{abcd} \psi_a^{(1)}\psi_b^{(2)}\psi_c^{(3)}\psi_d^{(4)}$$

Where $G^{abcd}$ is a hypermatrix of $16 = 2^4$ complex valued components which represents the graviton interaction. The symmetry in this action is four copies of $SU(2)$ operating on the individual 2 dimensional vector spaces corresponding to the four indices. Taking a leaf from the book of quantum information theory we use the fact that the complexification $SU(2,\mathbb{C})$ is isomorphic to $SL(2,\mathbb{C})$, so the overall symmetry can be regarded as $SL(2,\mathbb{C})^{\otimes 4}$.

Now we want to construct the partition function for this action which is given by

$$Z_q = \int e^{i\beta S_I} d^8\psi$$

If there were only two qubits the interaction would be represented by a 2x2 matrix instead of the 2x2x2x2 hypermatrix $G^{abcd}$. From basic quantum theory we know that the integration could then be done to give the inverse of the determinant of the matrix. With a four qubits interaction the integration is more difficult, but the work of Gelfand, Kapranov, and Zelevinsky on generalised hypergeometric functions [10] tells us that the leading terms to order $\beta^{-1}$ in such an integration is given by the inverse of a quantity known as the hyperdeterminant of $G^{abcd}$.

$$Z_q = \left(Det(\beta G)\right)^{-1} + O(\beta^{-2})$$

This hyperdeterminant is a degree 24 polynomial in the sixteen components of $G^{abcd}$ which can be constructed as an invariant of the $SL(2,\mathbb{C})^{\otimes 4}$ action on the hypermatrix. It was first studied in the 19[th] century by the algebraist Schläfli after Cayley described the simpler hyperdeterminant for a three qubit hypermatrix.

When fully expanded, Schläfli's hyperdeterminant is an expression with 2,894,276 terms [11] making it a bit hard to work with directly. Fortunately there is nevertheless a way to understand the meaning of this complicated result. That is because the hypermatrix $G^{abcd}$ describes an elliptic curve [12] which can be written in the form

$$G^{abcd}\psi_b^{(1)}\psi_c^{(2)}\psi_d^{(3)} = 0$$

This can be solved in steps by first writing it as a matrix equation

$$M^{ab}\psi_b^{(1)} = 0$$

$$M^{ab} = G^{abcd}\psi_c^{(2)}\psi_d^{(3)}$$

A solution to the equation exists iff the matrix is singular, i.e

$$\det(M) = 0$$

In the second step we regard this as an equation of a 2x2x2 hypermatrix

$$M^{ab} = A^{abc}\psi_c^{(2)}$$

$$A^{abc} = G^{abcd}\psi_d^{(3)}$$

The determinant of $M$ is then a quadratic in the components of $\psi_c^{(2)}$ and its solutions are determined by the square roots of the discriminant of the quadratic. This discriminant turns out to be none other than Cayley's hyperdeterminant for the hypermatrix $A$. So we now reduce the problem to the equation

$$y^2 = Det(A)$$

As a final step we set the components of the state $\psi_d^{(3)}$ as a vector $(x, 1)$. Cayley's hyperdeterminant is a polynomial of degree 4 so the equation now takes the form a homogeneous quartic

$$y^2 = Ax^4 + Bx^3 + Cx^2 + Dx + E$$

The coefficients $A, B, C, D, E$ are themselves fourth degree polynomials in the components of the hypermatrix $G$. Equations of this form are called Jacobi quartics and are in the same class as elliptic curves which take a similar form with a cubic on the left. In fact, given one solution of this equation there is a standard way to reduce the problem to the cubic elliptic form using elementary rational transformations. The textbook way to understand elliptic equations is to transform them into differential equations with $x$ a function of a complex variable $z$ and $y = \frac{dx}{dz}$.

$$\left(\frac{dx}{dz}\right)^2 = Ax^4 + Bx^3 + Cx^2 + Dx + E$$

The solutions to this equation are elliptic functions which are doubly periodic in the variable $z$. This in turn means that they can be regarded as functions on a Riemann surface in the form of a torus, i.e the complex plane is made periodic in both directions, but the way it joins up in the complex plane is determined by an offset parameter $\tau$ which depends on the coefficients of the Jacobi quartic.

Using the theory of elliptic curves it can be shown that the discriminant of the quartic $\Delta(\tau)$ as a function of $\tau$ is a well known modular form called Ramanujan's tau function which in turn is given by the 24[th] power of the Dedikind eta function

$$\Delta(\tau) = (\sqrt{2\pi}\, \eta(\tau))^{24}$$

This is highly significant because the one loop partition function for a bosonic string in 26 dimensions is [13]

$$Z_s = \frac{1}{\Delta(\tau)}$$

But the discriminant $\Delta(\tau)$ as a function of the components of the hypermatrix $G^{abcd}$ is equal to Schläfli's hyperdeterminant. Thus we arrive at the conclusion that the partition function for the 4-qubit system to first order is the same as the partition function of the bosonic string in 26 dimensions. This partition function counts all the vibration modes of the string world sheet which forms the two dimensional complex plane moving in the remaining 24 dimensions of space.

The remarkable conclusion is that a system of four qubits defined in the absence of any background spacetime is to first order equivalent to the 26 dimensional bosonic string theory with all 26 dimensions being emergent in the theory with perfect Lorentz invariance.

## The Holographic Principle and Complete Symmetry

We think of the universe as a vessel filled with matter and energy. The vessel is spacetime which may itself be emergent, but what about the material world, could that be emergent too? Is it possible that everything is emergent from the interaction of qubits of pure information?

Clearly four qubits is not sufficient to describe the universe, but a much larger number of them could be. The holographic principle tells us that the amount of information in space is not limited by a maximum density in the *volume* as you might expect. Instead it is limited by a maximum density on the *surface* that surrounds it. If the amount of information reaches the limit, then a black hole will form with the information stored on the surface of the event horizon. This means that you can have a high density of information in a small region, but the average density throughout the universe must be much lower.

We know that the holographic principle can be realised because Maldacena's CFT/AdS duality from string theory describes a 5 dimensional spacetime theory with gravitation in terms of a 4 dimensional description on the AdS horizon [14], but even in this case we do not know exactly how the mechanism works. In my opinion, the only way such holography can arise is if there is a hidden symmetry principle so that an application of gauge fixing removes all degrees of freedom in the higher dimensional space leaving just a description of a reduced theory on the boundary. This can only work if there is at least one degree of symmetry for every dynamic degree of freedom. In group theory language, the fundamental fields must be in a single representation of the symmetry which is no bigger than the adjoint representation. Most likely this would mean that precisely the adjoint representation is used. I call this circumstance "complete symmetry".

To understand what this means, think about how gauge fixing works in a pure gauge Yang Mills theory. The gauge field is described by a vector of component fields each of which is in the adjoint representation of the underlying gauge group. This means that in four dimensional spacetime there are exactly four times as many component fields as there are degrees of symmetry. Gauge fixing will only remove one quarter of them. In a holographic theory the amount of symmetry must be larger so that all the degrees of freedom in the bulk of space can be gauge-fixed away leaving just boundary field invariants. The process is analogous to diagonalising a matrix where the only remaining degrees of freedom are the eigenvalues on the diagonal.

The implication is that the full symmetry of nature is a much bigger symmetry than the ones we know. Even the grand unified theories and supersymmetric theories of gravity do not have complete symmetry in this sense. The symmetry principle of string theories is not understood but it is possible that an unknown hidden universal symmetry exists that can explain the holographic principle.

## The Foundations of String Theory

String Theory and its offshoots such as M-Theory are the most advanced theories we have for unification of all the physical forces and matter including gravity. But string theory is not without its problems. It seems to have a very large number of possible vacua with no criterion we can use for selecting the real one. However, we cannot be certain of this until we understand the foundations of string theory better. It is possible that mechanisms exist for tunnelling between vacua, making them unstable. Only a rigorous and non-perturbative formulation of string theory can determine this for us.

As we currently understand it, superstring theory is described by a range of different 10 dimensional versions with a perturbation series for each. These versions are unified through some non-perturbative duality transformations which suggest that they are really all part of the same M-theory in 11 dimensions. The trouble with this formulation is that it is incomplete and breaks down when we try to apply it to physics at Planck scale temperatures; just the regime where it should be most interesting. To resolve this, a non-perturbative formulation for the complete theory is required.

For a while it was thought that a matrix model [15] could fulfil this need, but the matrix theories are also incomplete because they do not allow a flexible choice of vacuum. As well as being non-perturbative, their major strength is that spacetime is emergent from the matrix model, just as we would like it to be. Matrix theories suggest that a purely algebraic formulation of M-theory with emergent spacetime may be possible.

To find a unified formulation for string theory we could look to the algebra of qubits, which is the algebra of hypermatrices and hyperdeterminants first studied in the nineteenth century. A typical hypermatrix is a multidimensional array of $d^n$ components. For qubits we would look at just the case $d = 2$. The symmetry group is $SL(d, \mathbb{C})^{\otimes n}$ which has $k = \frac{1}{2}d(d+1)n$ degrees of symmetry. For $n > 3$ and $d > 1$ the number of components is greater than the dimension of the symmetry $d^n > k$. This implies that complete symmetry is not possible using just qubits and local symmetry transformations.

But that is not the end of the story. In 1994 I described an "event-symmetric" algebra for string theory modelled as discrete bits linked in chains [16]. The algebra is in the form of a necklace super-Lie algebra that acts as both the operator algebra for a string field theory and it's algebra of symmetry generators. In this model the principle of complete symmetry is therefore realised.

The earliest suggestions that the universe runs on quantised bits came from Carl von Weizsäcker's ur-theory [17]. Weizsäcker had the idea that you could start from a single quantum bit and quantise it iteratively to build up all structures of physics. Wouldn't it be wonderful if the universe worked like that? In the event-symmetric theory of strings a similar principle of multiple quantisations is used to build up an algebraic structure for higher dimensional entities. This reflects Michael Green's earlier research on worldsheets for worldsheets [18]

The remaining challenge is to show that spacetime and the known forms of string theory can emerge from this algebra. This remains unsolved but the discovery that bosonic string theory could emerge from the interactions of qubits using elliptic curves, as described above, provides hope that this could now be possible.

**Information Physics and Technology**

When Shannon began exploring the links between digital information and entropy he could barely have imagined the remarkable application his ideas now have in today's technology. Data compression algorithms which depend on the understanding he provided are used in millions of video and audio gadgets sold around the world. When information is transmitted over the internet or telecommunication systems it must be expanded using mathematical error correcting codes that pad the signal with redundant data so that the original signal can be efficiently reconstructed by the receiver, even if some random bits are lost during the transmission. Shannon's theory of information content and entropy are crucial to making such technology work.

While these information technologies are currently used in a plethora of new electronic devices, a future generation of possibilities is being studied by mathematicians and physicists. This will eventually lead to the age of quantum computers, so powerful that we can scarcely imagine their capabilities and applications. But they may never realise their promise unless we can solve some fundamental problems.

The qubits in a quantum computer can store much more information than the classical bits in existing computers. The entangled quantum states of qubits can be a supposition of many possible bit states at once and a quantum computer should be able to perform many computations in parallel using this information as data. The computational power of a quantum computer should in theory increase exponentially with the number of qubits it uses. So far quantum computers have been constructed with just a few qubits. In time more useful systems will be developed but they will only realise their full potential if the logical operations they perform avoid the problems of quantum decoherence that threaten to spoil any calculation. Even if the quantum computer is kept maximally isolated from outside influence, there will be a limit to how long it can run before the state is changed by interference from outside.

One solution that quantum information engineers hope to use against the decoherence problem is to apply quantum error correcting codes to keep the information sound. This would work in much the same way as error correcting codes that correct data losses in communications, but it would have to work continuously during any calculation process.

If the universe works as a quantum computer you may ask yourself, how does it avoid the same problems of decoherence? At this stage we don't know the answer or even if the question defines a real problem, but already physicists have found that error correcting codes arise naturally in the workings of string theory. A particular code known is a Hamming code which emerges when branes from M-theory are wrapped around the compactified dimensions of spacetime [19,20]. This wrapping can only be done according to certain combinatorial rules. The result is found to represent qubits of information according to whether a brane is wrapped or not, and the solutions to the rules when the compactified space has the simplest possible topology produce bit patterns that are identical to the Hamming code bit patterns.

If string theory is correct the actual compactification needed to reduce it to the four dimensional theory we know, may use a more complicated Calabi-Yau manifold. This is a geometrical construction that can be viewed as a generalisation of an elliptic curve. It is therefore significant that elliptic curves are related to the invariants of qubit systems as I have described above. It may be that nature has chosen a Calabi-Yau manifold that not only reproduces the standard model of physics which is ideal for supporting life, but it also generates an error correction system that controls the stability of the universe so that it does not melt into a messy quantum supposition of states. In effect, the Calabi-Yau space and the fluxes that define our vacuum are the program of the quantum computer that runs our universe.

Of course this is currently no more than speculation inspired by some interesting observations that relate string theory and quantum information theory, but if we do not pursue such ideas we may miss the chance to understand how to develop the next generation of computer technology.

## Conclusion

Is reality digital or analogue? The universe may run like a quantum computer processing qubits of information, but a quantum computer is neither a digital computer nor an analogue computer in the traditional sense of those terms. Qubits are based on binary information but the quantisation means that states can be in a supposition of true and false values, with a complex number valued amplitude. You might say that the complex values turn it into an analogue computer, but it's not that simple. Individual amplitudes cannot be measured, only statistical averages, and a quantum computer does not actually work by taking such averages.

What about space, time and matter, are they discrete or continuous? Again the answer is open to interpretation. Space and time could emerge from interactions between discrete entities, yet their symmetries are continuous and perfect. They are not mere approximations to a discrete underlying process. The amount of information in a given region of space is finite. You cannot measure unlimited detail down to any length scale, yet this does not mean that it is build from discrete units. It is sometimes said that spacetime at the Planck scale should be like a bubbling foam, but string theory suggests a smooth structure at least at cold temperatures and now the Fermi gamma ray observations back this up. At temperatures approaching the Planck scale and beyond, the situation is very different. Smooth space will break down and time will cease to have a meaning at the not regimes present in the big bang, or in the singularities at the centre of black holes.

Reality then is neither digital nor analogue and yet it is both. In the strange world of the quantum these classical ideas are not good enough for the job of interpreting the way the universe runs. Only the rules of quantum theory can adequately describe the strange universe in which we live.